\documentstyle[12pt]{article}
\newcommand{\beq}{\begin{equation}}
\newcommand{\eeq}{\end{equation}}
\newcommand{\beqa}{\begin{eqnarray}}
\newcommand{\eeqa}{\end{eqnarray}}
\newcommand{\beqar}{\begin{eqnarray*}}
\newcommand{\eeqar}{\end{eqnarray*}}

\newcommand{\al}{\alpha}

\newcommand{\k}{\kappa}

\newcommand{\cA}{{\cal A}}

\newcommand{\eg}{{\it e.g.,}\ }
\newcommand{\ie}{{\it i.e.,}\ }

\newcommand{\norm}[1]{\raise.3ex\hbox{:}#1\raise.3ex\hbox{:}}

\newcommand{\labell}[1]{\label{#1}}
\newcommand{\labels}[1]{\label{#1}} 

\newcommand\e{{\rm e}}

\newcommand\prt{\partial}

\begin{document}
\begin{titlepage}
\rightline{\small hep-th/9703041 \hfill McGill/97-02}
\vskip 5em

\begin{center}
{\bf \huge
New angles on D-branes}
\vskip 3em

{\large J.C. Breckenridge\footnote{email:
jake@haydn.physics.mcgill.ca\hfil}},
{\large G. Michaud\footnote{email:
gmichaud@hep.physics.mcgill.ca\hfil}} and
{\large R.C. Myers\footnote{email: rcm@hep.physics.mcgill.ca\hfil}}
\vskip 1em

{\em	Department of Physics, McGill University \\
        Ernest Rutherford Physics Building\\
	Montr\'eal, Qu\'ebec, Canada H3A 2T8}
\vskip 4em

\begin{abstract}
A low-energy background field solution is presented which describes
several D-membranes oriented at angles with respect to one another.
The mass and charge densities for this configuration
are computed and found to saturate the BPS bound, implying
the preservation of one-quarter of the supersymmetries.  T-duality
is
exploited to construct new solutions with nontrivial angles
from the basic one.
\end{abstract}
\end{center}

\end{titlepage}

\setcounter{footnote}{0}
\section{Introduction}

Understanding of non-perturbative aspects of string theory has
advanced rapidly during the past two years \cite{drama}.
For example, all five consistent superstring theories can now
be related through the use of various string dualities.
These connections suggest that all of these string theories
are really perturbative expansions about different
points in the phase space of a more fundamental framework,
commonly called M-theory. The development of these string dualities
has brought with it the realization that extended objects
beyond simply strings
play a crucial role in these theories. In the case of the
Type II (and I) superstrings, of particular interest
are Dirichlet branes (D-branes) which
carry charges of the Ramond-Ramond (RR)
potentials\cite{Polchin}.

D-branes have also proven their worth from a calculational
standpoint. For example,
bound states of D-branes have recently facilitated the computation
of the entropy of black holes from a counting of
the underlying microscopic degrees of freedom\cite{blackholes}.
In these analyses, the bound state configurations must be
supersymmetric
(or nearly so --- see, however, \cite{atish}) in order to protect
the
counting of states from loop corrections by BPS saturation.
In this case, the microscopic counting which is made at weak
coupling
can be compared with the expected degeneracy of states at strong
coupling
at which the bound state has formed a black hole.
This is one of the reasons for which supersymmetric D-brane
configurations are of particular interest.

A great deal of effort has gone into generating the low-energy
background field solutions corresponding to various D-brane
bound states\cite{revbrane}. These solutions are restricted to those
describing p-branes which are either parallel or intersect
orthogonally. It has been shown\cite{Berkooz}, however,
that there exist supersymmetric configurations where
the angles between the D-branes are {\it other\/} than zero or
$\pi/2$.
Preserving supersymmetry in such multiple D-brane configurations
requires that the angles are restricted to lie in an $SU(N)$
subgroup
of rotations.
The corresponding background field configurations remain largely
unexplored, but in this paper, we will present one such class of
solutions. Our basic solution describes any number of D-membranes
whose relative orientations are given by certain $SU(2)$ rotations.

The paper is organized as follows: section 2 presents
our solution and calculation of the mass and charge densities
for this system of angled D-branes.  With the latter, we demonstrate
that the BPS bound saturated by this configuration.
In section 3, we exploit T-duality to create
solutions involving angled D3- and D4-branes, as well as some more
exotic
configurations, arrived at by considering T-duality along
world-volume
coordinates of D-membranes in our original solution. Finally,
a brief discussion follows in section~4. Our notation
and conventions follow those established in \cite{us}.

\newpage
\section{Membranes at angles} \labels{angles}

We begin by writing down the  solution describing an arbitrary
number $n$ of D-membranes, each of which is rotated by certain
$SU(2)$ angle, in the type IIA low energy effective
string theory. The solution contains only a nontrivial
(string-frame) metric, three-form RR potential and dilaton:
\beqa
ds^2 & = & \sqrt{1 + X} \Bigg [ {1 \over 1+ X} \bigg( - dt^2
+\sum_{j=1}^4(d y^j)^2\nonumber\\
& & \qquad \qquad \quad
+ \sum_{a=1}^n X_a\,\bigg \{ \big[{(R_a)^1}_i d y^i\big]^2 +
\big[{(R_a)^3}_j d y^j\big]^2\bigg \} + \sum_{i=5}^9 (d x^i)^2 \Bigg
]
\nonumber\\
A^{(3)} & = & {dt \over 1+X} \wedge \bigg \{ \sum_{a=1}^n X_a\
(R_a)^2{}_i
d y^i \wedge (R_a)^4{}_j d y^j \nonumber\\
& & \qquad \qquad \quad
- \sum_{a < b}^n X_a X_b\,\sin^2 (\alpha_a - \alpha_b)\
(d y^1 \wedge d y^3 - d y^2 \wedge d y^4) \bigg \}\nonumber\\
e^{2 \phi_a} & = & \sqrt{1 + X}
\labell{newsol}
\eeqa
where
\beq
X = \sum_{a=1}^n X_a + \sum_{a< b}^n X_a X_b\, \sin^2( \alpha_a -
\alpha_b)\ \ .
\eeq
Above, the rotation matrix $R_a$ associated with the $a$'th
D-membrane is given by
\beq
R_a = \left(\begin{array}{cc}
	{\begin{array}[t]{cr} \cos \alpha_a & -\sin \alpha_a \\
				\sin \alpha_a &\cos \alpha_a
\end{array}} &
				{\raisebox{-10pt}{\rm\huge 0}}\\
				{\raisebox{-6pt}{\rm\huge 0}} &
{\begin{array}{rc}
				\cos \alpha_a & \sin \alpha_a \\
				- \sin \alpha_a & \cos \alpha_a
\end{array}}
	\end{array} \right)
\labell{rotate}
\eeq
The matrices acting in the space of $y^i$'s
are easily recognized as $SU(2)$ rotations as follows:
one defines the complex coordinates $z^1=y^1+iy^2$ and
$z^2=y^3+iy^4$.
Then the above rotations are given by $(z^1,z^2)\rightarrow
(e^{i\alpha_a}z^1,e^{-i\alpha_a}z^2)$, or $z^i\rightarrow
\left[\exp(
i\alpha_a\sigma_3)\right]^i{}_j z^j$.
One expects from \cite{Berkooz} that restricting
the relative orientation of the membranes in this way will
preserve some of the supersymmetry, and we confirm this fact in
the following.

The functions $ X_a$ are harmonic functions in the transverse space
of $x^i$'s. That is, they solve the flat-space Poisson's equation
in the transverse space, \eg
\beq
\delta^{ij}\prt_i\prt_j X_a =
-\ell_a^3\cA_4\,\prod_{k=5}^9\!\delta(x^k-x^k_a)
\ \ .
\labell{laplace}
\eeq
yielding the solutions
\beq
X_a(\vec x) =\, { 1\over 3}
\bigg( {\ell_a \over \vert \vec x - \vec x_a\vert}\bigg)^3\ \ .
\labell{laplacesol}
\eeq
Above, $\ell_a$ are arbitrary
positive parameters which have the dimension of length,
and we use $\cA_4$ to denote the volume of a unit four-sphere.
In general, one has $\cA_{n-1}=2\pi^{n/2}/\Gamma(n/2)$.
In fact, one may introduce any number of delta-function sources at
arbitrary positions on the right hand side of eq.~(\ref{laplace}),
and the corresponding solution would describe a system of parallel
branes.

A few words are in order as to the origin of this solution.  It is
in
effect an interpolation between the known solutions for
parallel D-membranes, and that\cite{Tseytlin} for orthogonal
D-membranes
intersecting over a point.  It is straightforward to verify that
when the angles are all set to $\alpha_a = 0$, the solution reduces
to that of $n$ parallel branes lying in the ($y^2,y^4$) plane.
Note that in this case the membranes have also been delocalized or
smeared out in the $y^1$ and $y^3$ directions.
One may also verify that choosing all $\alpha_a=\alpha_o$ simply
corresponds to an overall $SU(2)$ rotation of the previous solution.
Similarly the known configuration of orthogonally oriented membranes
is reproduced by choosing $ \alpha_a$'s to be either zero or
$\pi/2$.
Further with the
$\alpha_a$ set to either $\alpha_o$ and $\pi/2+\al_o$,
eq.~(\ref{newsol})
corresponds to a rotation of this solution. Finally, one may
verify that making a further $SU(2)$ rotation of the entire solution
simply corresponds shifting all of the angles $\alpha_a$ by the same
constant. For this to work, it is important that the second term
in $A^{(3)}$ is proportional to $dt\wedge {\rm Re}(dz^1\wedge
dz^2)$, which is
invariant under $SU(2)$ rotations.
Verifying that eq.~(\ref{newsol}) solves the
low-energy field equations of type IIA string theory was
only done with the aid of a computer.

One final comment on our notation:
we refer to $x^i$ and $y^i$ as transverse and
world-volume coordinates, respectively. For a given brane, however,
a
particular (linear combination of) $y^i$ may actually still
correspond
to a transverse direction, although it will be one in which the
brane
is delocalized. Hence in the next section, when we smear out
the solution in some $x^i$ making the solution independent of this
coordinate, the designation for the coordinate is changed to $y^i$.
We will also assume
that the $y^i$ coordinates are all compact with a range of $2\pi
L_i$

\subsection{Mass and Charge Relations} \labels{cquant}

In this section, we consider some of the physical characteristics of
the above configuration (\ref{newsol}). In particular, we calculate
the mass and charge densities of our solution. The latter densities
are calculated using asymptotic flux integrals, and so they are
completely determined by the leading-order
behavior of the asymptotic fields. In examining the solution,
one sees that these leading order fields are simply linear
superpositions
of the asymptotic fields generated by the individual rotated
membranes.
Hence we generalize the rotation appearing in these linearized
fields
by replacing $\alpha_a$ by an independent angle $\beta_a$ in the
lower two-by-two block of the rotation matrices (\ref{rotate}).
Such a configuration would only solve the linearized asymptotic
equations
of motion, and not the full nonlinear supergravity equations, but
this
generalization does yield some interesting insight when examining
the
BPS bound.

For a $p$-brane, the ADM mass per unit
$p$-volume is defined as\cite{massy}:
\beq
m ={1\over 2\k^2}\oint \sum^{9-p}_{i=1} n^i\left[
\sum^{9-p}_{j=1} \left(\partial_j h_{ij} -\partial_i h_{jj}\right)
-\sum^{p}_{a=1} \partial_i h_{aa}\right] r^{8-p} d\Omega
\labell{mform}
\eeq
where $n^i$ is a radial unit vector in the transverse space and
$h_{\mu\nu}$ is deformation of the {\it Einstein-frame} metric
\beq
h_{\mu\nu} = g^E_{\mu\nu} -\eta_{\mu\nu}
\labell{hhh}
\eeq
from flat space in the asymptotic region.
Calculating the mass per unit four-volume (of the internal space of
$y^i$'s)
for our angled system by means of (\ref{mform})
gives us the result
\beq
m = {{\cal A}_4 \over 2 \kappa^2} \sum_{a=1}^n \ell_a^3\ \ .
\labell{mass}
\eeq
Thus the mass
density is simply the sum of the mass densities of the constituent
branes, which was to be entirely expected. Note then that
this result is completely independent of the rotation angles.

The membranes carry an electric RR four-form field strength and
the corresponding physical charge density is
given by\cite{report}
\beq
q = {1\over\sqrt{2}\k} \oint{}^* F^{(4)}\ \ .
\labell{chargedefs}
\eeq
Hodge duality produces a six-form which is then integrated
over the asymptotic four-sphere in the transverse space and
some two-torus in ($y^1,y^2,y^3,y^4$). Thus given,
the three-form potential in eq.~(\ref{newsol}), in applying
(\ref{chargedefs}) we obtain a number of independent charges
related to the choice of asymptotic surface over which one
integrates. For example the term in $A^{(3)}$ proportional
$dt\wedge dy^2\wedge d y^4$ yields a term in ${}^* F^{(4)}$
to be integrated over the compact
coordinates $y^1$ and $y^3$ as well as the four-sphere
at infinity.  We use the following notation to write
the resulting charge
\beq
q_{13} = -q_{31}= {{\cal A}_4 \over \sqrt{2} \kappa}
(4 \pi^2 L_{1}L_{3})\sum_{a=1}^n \mu_a\ell_a^3
\cos \alpha_a\, \cos \beta_a
\labell{qxxcharge}
\eeq
where the antisymmetric matrix notation will be useful later on.
This result gives the charge per unit area in the
($y^2,y^4$) plane, \ie the plane in which the branes lie for
$\alpha_a=\beta_a=0$. In order to compare the charges, however,
we should divide out the area of the orthogonal ($y^1,y^3$) torus
in order
to produce a charge per four-volume in the entire compact space.
Hence we define $\tilde q_{13} = q_{13}/(4\pi^2 L_1L_3)$.
In a like manner all the charge densities $\tilde  q_{ij}$
can be calculated and we list the nonvanishing contributions
\beqa
\tilde q_{13} & = & -{{\cal A}_4 \over \sqrt{2} \kappa} \sum_{a=1}^n
\ell_a^3 \cos \alpha_a\, \cos \beta_a \nonumber\\
\tilde q_{14} & = & -{{\cal A}_4 \over \sqrt{2} \kappa} \sum_{a=1}^n
\ell_a^3  \cos \alpha_a\,\sin \beta_a \nonumber\\
\tilde q_{23} & = & {{\cal A}_4 \over \sqrt{2} \kappa} \sum_{a=1}^n
\ell_a^3 \sin \alpha_a\, \cos \beta_a \nonumber\\
\tilde q_{24} & = &  {{\cal A}_4 \over \sqrt{2} \kappa} \sum_{a=1}^n
\ell_a^3 \sin \alpha_a \sin \beta_a\,.
\labell{charges}
\eeqa
Of course these charge densities are dependent on the
rotation angles which orient the various D-membranes.
Note that if $ \alpha_a = \beta_a = 0$ we recover the expected
charge
configuration of a collection of parallel membranes lying in
the ($y^2,y^4$) plane, \ie
\beq
\tilde q_{13}  = -{{\cal A}_4 \over \sqrt{2} \kappa} \sum_{a=1}^n
\ell_a^3\qquad
\tilde q_{14} =  \tilde q_{23}
= \tilde q_{24} = 0
\eeq
where the single nonvanishing charge density is simply the sum of
that
for the individual branes.

Having calculated these physical characteristics of our
configuration
of D-membranes with angles, we would like to examine the BPS bound.
The latter may be determined from the eigenvalues of the
Bogomol'nyi matrix, which is derived using both the supersymmetry
algebra and the asymptotic form of the background
fields\cite{older}.
Unbroken supersymmetries arise when this matrix has eigenspinors
with a vanishing eigenvalue.
In the present problem, the Bogomol'nyi matrix is\cite{Popelu}
\beq
{\cal M} = m + {1 \over \sqrt{2}\kappa} \tilde q_{ij} \Gamma_{0 i j}
\labell{bpsone}
\eeq
for which the distinct eigenvalues are
\beq
m \pm {1 \over \sqrt{2} \kappa} \sqrt{\tilde q_{ij}
\tilde q_{ij}
\pm {1 \over 2} \epsilon_{ijkl}\tilde q_{ij}\tilde q_{kl}\,\,}\, .
\labell{bpseigen}
\eeq
In these formulae, the implicit sums all run from 1 to 4, and we use
the antisymmetric notation $\tilde q_{ij}=-\tilde q_{ji}$
introduced above.
Also note that the two signs in the eigenvalues are chosen
independently.
Since the mass is positive, the eigenvalues for which the first sign
is positive cannot vanish, and hence at least half of the
supersymmetries
are broken by our solution. The vanishing of the remaining
eigenvalues
can be expressed in terms of a BPS mass limit
\beq
m_{\pm}^2={1 \over {2} \kappa^2} \left(\tilde q_{ij}
\tilde q_{ij}
\pm {1 \over 2} \epsilon_{ijkl}\tilde q_{ij}\tilde q_{kl}\right)\ \
{}.
\labell{bownd}
\eeq
Substituting our values for the charge densities (\ref{charges})
results
in
\beq
m_{\pm}^2= \left({{\cal A}_4 \over {2} \kappa^2}\right)^2
\left[\left(\sum_{a=1}^n \ell_a^3 \cos( \alpha_a \mp
\beta_a)\right)^2
 + \left(\sum_{a=1}^n \ell_a^3 \sin( \alpha_a \mp \beta_a)
\right)^2\right]
\ .
\labell{bpstwo}
\eeq
In comparing these BPS bounds (\ref{bpstwo}) with the mass
(\ref{mass}),
we find that in general the mass exceeds the former bounds. To make
this
more apparent, one may introduce complex variables $Z_{\pm,a}=
({\cal A}_4/2\kappa^2)\ell_a^3\,\exp[i(\alpha_a \mp \beta_a)]$.
Now it is clear that generically $m^2=(\sum_a|Z_{\pm,a}|)^2$ exceeds
$m_\pm^2=|\sum_a Z_{\pm,a}|^2$. It is also clear that the
only way to lower the mass to one of the bounds is to chose all of
the phases
to be equal, \ie $\alpha_a-\beta_a=2\theta$ or
$\alpha_a+\beta_a=2\theta'$.
There are only two distinct choices here up to an overall rotation.
If we set $\alpha_1=\beta_1=0$ to fix the overall orientation of the
configuration, we must choose the remaining angles with
$\beta_a=\alpha_a$
or $\beta_a=-\alpha_a$. The former corresponds to the choice
made in our solution (\ref{newsol}),
and for which we then have $m=m_+$ and one-quarter
of the supersymmetries being preserved.  The latter
choice, for which $m=m_-$,
would yield a slightly different configuration. Complex
$SU(2)$ rotations are again relevant in this case, but now the
$SU(2)$ acts on ($z^1,\bar{z}^2$)=($y^1+iy^2,y^3-iy^4$). Our
solution
would be modified by changing the sign of $\alpha_a$ in the lower
two-by-two
block of the rotation matrices (\ref{rotate}), and the sign of
$dy^2\wedge dy^4$ would be reversed in the last term in $A^{(3)}$.
As expected, our results here are entirely consistent with
the analysis of \cite{Berkooz} mentioned earlier which is formulated
at the level of the string world-sheet and provide
an independent confirmation of their results when applied to
D-membranes.

\section{T-Duality} \labels{tdual}

The ten-dimensional T-duality map between the type IIA and IIB
string theories was given in ref.~\cite{Ortin} --- see
\cite{us} for the transformation using the present conventions.
In the next subsection,
we consider the effect of T-duality along coordinates that are
in the transverse space. The effect of these transformations is
to extend the dimension of the D-branes. The results then are
new solutions describing D$p$-branes with relative $SU(2)$ angles
and so remaining parallel over a ($p$-$2$)-brane.
In subsection \ref{worldvol},
we consider the effect of T-duality transformations along
world-volume
coordinates. The results here involve more exotic bound state
configurations of D-branes, as found in \cite{us}.

\subsection{Transverse directions}\labels{transverse}

In order to apply T-duality along one of the transverse coordinates,
\eg $x^5$, we must first delocalize the solution in this direction,
which we then denote as $y^5$. This amounts to replacing the
sources in eq.~(\ref{laplace}) by four-dimensional delta-functions,
producing solutions of the form
\beq
X_a(\vec x) =\, { 1\over 2}
\bigg( {\ell_a \over \vert \vec x - \vec x_a\vert}\bigg)^2
\eeq
where now $\vec x=(x^6,x^7,x^8,x^9)$.
A straightforward application of the T-duality map from the type IIA
to the type IIB theory  along $y^5$ in this smeared out solution
yields
\newpage
\beqa
ds^2 & = & \sqrt{1 + X} \Bigg [ {1 \over 1+ X} \bigg( - dt^2
+\sum_{i=1}^5(d y^i)^2 \nonumber\\
& & \qquad \qquad \quad
+ \sum_{a=1}^nX_a\, \bigg \{ \big[{(R_a)^1}_i d y^i\big]^2 +
\big[{(R_a)^3}_j d y^j\big]^2\bigg \} + \sum_{i=6}^9 (d x^i)^2 \Bigg
]
\nonumber\\
F^{(5)} & = & dt\wedge d y^5\wedge
dx^k\wedge\partial_k \Bigg \{{1 \over 1+X}\bigg [
\sum_{a=1}^n X_a\ {(R_a)^2}_i d y^i \wedge {(R_a)^4}_j d y^j
\nonumber\\
& & \qquad \qquad
- \sum_{a < b}^n X_a X_b\ \sin^2 (\alpha_a - \alpha_b)\
(d y^1 \wedge d y^3 - d y^2 \wedge d y^4) \bigg ]\Bigg \}\nonumber\\
& & \qquad \qquad \mbox{} +
dx^h\wedge dx^i\wedge dx^j
\wedge\, \epsilon_{hijk} \partial_k \Bigg \{
\sum_{a=1}^n X_a\ {(R_a)^1}_l d y^l \wedge {(R_a)^3}_m d y^m\Bigg\}
\nonumber\\
e^{2 \phi_b} & = & 1\  .
\labell{threesec}
\eeqa
This solution obviously describes a system of angled D3-branes,
as indicated by the presence of the nontrivial five-form RR
field strength. We have written the solution in terms of the
self-dual field strength, rather than the potential $A^{(4)}$,
because the magnetic part of the latter is rather unwieldy
when the D3-branes are centered at arbitrary positions $\vec x_a$.
If one sets $\vec x_a=0$, the potential can be given in a fairly
compact form using polar coordinates on the transverse space.
Note also that $\epsilon_{hijk}$ is the antisymmetric Levi-Civita
symbol on the transverse space with $h,i,j,k = 6\dots 9$
and $\epsilon_{6789}=+1$.

One can carry this process further by delocalizing the above
solution
in another transverse
coordinate $x^6$ (which we then denote $y^6$ --- also, note
that one now has $X_a=\ell_a/|\vec x-\vec x_a|$), and applying
T-duality
along this direction to produce a system of D4-branes with $SU(2)$
angles.
Here, the T-duality map
from type IIB to type IIA generates a
magnetic three-form potential
through $A^{(3)}_{ \mu \nu \rho} = A^{(4)}_{ \mu \nu \rho 6}$
(the remaining terms in this relation vanish in the present case).
This part of the transformation is equivalent to mapping the field
strengths
$F^{(4)}_{ \mu \nu \rho \sigma} = F^{(5)}_{ \mu \nu \rho \sigma 6}$,
since the delocalized solution is independent of $y^6$.
Hence the T-dual solution may be expressed as
\beqa
ds^2 & = & \sqrt{1 + X} \Bigg [ {1 \over 1+ X} \bigg( - dt^2
+\sum_{i=1}^6(d y^i)^2\nonumber\\
& & \qquad \qquad \quad
+ \sum_{a=1}^nX_a \bigg \{ \bigg[{(R_a)^1}_j d y^j\bigg]^2 +
\bigg[{(R_a)^3}_j d y^j\bigg]^2\bigg \} + \sum_{i=7}^9 (d x^i)^2
\Bigg ]
\nonumber\\
F^{(4)} & = &
dx^i\wedge dx^j
\wedge\, \epsilon_{ijk} \partial_k \Bigg \{
\sum_{a=1}^n X_a\ {(R_a)^1}_l d y^l \wedge {(R_a)^3}_m d y^m\Bigg\}
\nonumber\\
e^{2 \phi_b} & = & {1 \over \sqrt{1 + X}}\, .
\labell{foursec}
\eeqa
Again the magnetic field strength takes a much more compact form
than
the corresponding potential for the multi-center solution.
One sees that this solution obviously describes a system
of D4-branes since the magnetic $F^{(4)}$ is the only
nontrivial RR field.

Of course, this procedure of T-dualizing in the transverse space
can be continued to produce
configurations of higher dimensional D-branes with angles.
Since the $SU(2)$ rotations effectively extend the dimension of
the world-volume by two, the remaining solutions will have a
transverse space of dimension lower than three, and hence will not
be
asymptotically flat. For example, the solution describing angled
D6-branes
would have a transverse space of dimension one, and thus would have
the appearance of an anisotropic domain wall.

\subsection{World-volume directions} \labels{worldvol}

An alternative to the above procedure is to apply
T-duality in the world volume directions of the original
solution (\ref{newsol}). Since the membranes are rotated
in these directions, T-dual configurations will involve
D-brane bound states for which the difference in dimension is two,
as discussed in \cite{us}. To simplify the procedure
we specialize the general solution to the case
of two D-membranes and also set the rotation angles
$(\alpha_1,\alpha_2)=(0,
\alpha)$. With these simplifications, eq.~(\ref{newsol}) reduces to
\beqa
ds^2 & = &\, \sqrt{1+X}\Bigg\{{1 \over 1+X}\bigg(
- dt^2 + (1+X_1)\big[(d y^1)^2 + (d y^3)^2\big] + (d  y^2)^2 + (d
y^4)^2
  \nonumber\\
& &\qquad \qquad\mbox{}
+ X_2\left[( \cos \alpha d y^1 - \sin \alpha d y^2)^2
+ ( \cos \alpha d y^3 + \sin \alpha d y^4)^2\right]\bigg)
+ \sum_{i=5}^9 (d x^i)^2\Bigg\} \nonumber\\
A^{(3)} & = &\, {dt \over 1+X} \wedge\bigg\{
- (X_2+ X_1 X_2)\sin^2\alpha\,
d y^1 \wedge d y^3  + X_2 \sin\alpha\cos \alpha\,
d y^1 \wedge d y^4 \nonumber\\
& & \qquad \qquad\mbox{}
 - X_2 \cos\alpha\sin \alpha\, d  y^2 \wedge d y^3
+ (X_1 + X_2\cos^2\alpha +X_1 X_2 \sin^2 \alpha)\,  dy^2 \wedge dy^4
 \bigg\}\nonumber\\
\e^{2 \phi_a} & = &\, \sqrt{1 +X}
\labell{newsol2}
\eeqa
and $X$ is given by
\beq
X =\, X_1 + X_2 + X_1 X_2 \sin^2 \alpha .
\eeq
We also simplify the following results by positioning the second
membrane at the origin, \ie we set $\vec x_2=0$, but leave $\vec
x_1$
arbitrary.

As the first example, we apply T-duality
along the $y^4$ direction --- note that this
direction is tangent to the world-volume of the $a$=1 membrane, but
is angled with respect to the second.
We find
that
\beqa
ds^2 & = &\, \sqrt{1+X}\Bigg\{{1 \over 1+X}\bigg(
- dt^2 + (1 + X_1)(d y^1)^2 + (d  y^2)^2  \nonumber\\
& &\qquad \qquad\mbox{}
+ X_2( \cos \alpha d y^1 - \sin \alpha d y^2)^2
\bigg)+ {(d y^3)^2 + (d y^4)^2  \over 1 + X_2 \sin^2
\alpha}\nonumber\\
& & \qquad \qquad\mbox{} + d r^2
+ r^2( d \theta^2 + \sin^2 \theta( d \phi_1^2
+ \sin^2 \phi_1\,( d \phi_2^2 + \sin^2 \phi_2 d \phi_3^2)))\Bigg\}
\nonumber\\
A^{(4)} & = &\,  -{1 \over 2}X_2 \sin^2 \alpha \bigg \{
{1 + X_1 \over 1 +X} +{ 1 \over  1 + X_2 \sin^2 \alpha}\bigg \}
dt \wedge d y^1 \wedge d y^3 \wedge d y^4\nonumber\\
& & \qquad \qquad -{1 \over 2} X_2 \cos \alpha \sin \alpha
\bigg\{ {1 \over 1 + X_2 \sin^2 \alpha}+{ 1 \over 1 + X} \bigg \}
d t \wedge d y^2  \wedge d y^3 \wedge d y^4\nonumber\\
& &\qquad \qquad\mbox{}\! + \ell^3_2 \sin \alpha \sin^3 \theta
\sin^2 \phi_1\cos \phi_2\, ( \cos \alpha\,
d y^1 - \sin \alpha\, d y^2) \wedge d \theta
\wedge d \phi_1 \wedge d \phi_3\nonumber\\
A^{(2)} & = & {dt \over 1+ X} \wedge \bigg \{
X_2 \cos \alpha \sin \alpha d y^1 +
(X - X_2 \sin^2 \alpha) d y^2\bigg \} \nonumber\\
B^{(b)} & = & {X_2 \cos \alpha \sin \alpha \over 1 + X_2 \sin^2
\alpha}
d y^3 \wedge d y^4 \nonumber\\
\e^{2 \phi_b} & = &\, {1 + X \over 1 + X_2 \sin^2 \alpha}
\labell{partialone}
\eeqa
where we have transformed the coordinates transverse to the system
into spherical coordinates to facilitate the computations of the
four-form RR potential. Setting $X_2=0$, one can verify that this
solution reduces to that of a D-string lying parallel to $y^2$ and
at the same time delocalized in $y^1,$ $y^3$ and $y^4$. Setting
$X_1=0$
and comparing with the solutions of \cite{us}, one finds that the
solution is precisely that of a D(3,1)-brane bound state. There
has been a rotation of this bound state so that it lies in
($\cos\al\,y^2+\sin\al\,y^1,y^3,y^4$) with the D-strings oriented
along the
first direction. The bound state is also delocalized in the
orthogonal
$\cos\al\,y^1-\sin\al\,y^2$ direction.
The angle $\alpha$ also determines the relative charge densities of
the D-strings and D3-branes --- in \cite{us},
$\varphi=\pi/2-\alpha$.

Next we continue by applying T-duality in the $y^2$ direction
producing
a solution of the form
\beqa
d s^2 & = & \sqrt{1 + X} \bigg \{ {-dt^2 \over 1+X}
+ { (d y^1)^2 + (d y^2)^2 + (d y^3)^2 + (d y^4)^2 \over 1 + X_2
\sin^2 \alpha}
+ \sum_{i=5}^9 ( d x^i)^2 \bigg \} \nonumber\\
A^{(3)} & = & {X_2 \cos \alpha \sin \alpha \over 1 + X_2 \sin^2
\alpha}
dt \wedge ( d y^1 \wedge d y^2 - d y^3 \wedge d y^4 ) \nonumber\\
& & \qquad \qquad \mbox{}
+ \ell_2^3 \sin^2 \alpha\, \sin^3 \theta\, \sin^2 \phi_1\,
\cos \phi_2\, d \theta \wedge d \phi_1 \wedge d \phi_3 \nonumber\\
A^{(1)} & = & {\bigg \{ {1 + X_2 \sin^2 \alpha \over 1+ X} - 1
\bigg \}} dt
\nonumber\\
B^{(a)} & = & {X_2 \cos \alpha\, \sin \alpha \over 1 + X_2 \sin^2
\alpha}
( d y^3 \wedge d y^4 - d y^1 \wedge d y^2) \nonumber\\
e^{2 \phi_a} & = & {(1 + X)^{3 \over 2} \over (1 + X_2 \sin^2
\alpha)^2}.
\labell{partialtwo}
\eeqa
In this case setting $X_2=0$ reduces  the
solution to that of a D-particle positioned at $\vec x_1$ and
delocalized in the $y^i$ directions. Setting $X_1=0$
reproduces a special case of the D(4,2,2,0)-brane bound state given
in \cite{us}. Here the two angles of that solution are related,
\ie $\varphi=-\psi=\pi/2-\alpha$. The full solution then describes
the configuration conjectured by Lifschytz\cite{lifschytz} from
the consideration of the D-brane scattering processes.

As a final example, we perform T-duality along $y^3$ in the
two membrane solution (\ref{newsol2}) with the resulting solution
\beqa
ds^2 & = & \sqrt{1 + X} \Bigg \{ {-dt^2 +
(1 + X_1 + X_2 \cos^2 \alpha) (d y^1)^2 +
(1 + X_2 \sin^2 \alpha)(d y^2)^2 \over 1+X}\nonumber\\
& & \qquad \qquad\mbox{} - {2X_2 \cos \alpha \sin \alpha\,
d y^1 d y^2 \over 1+X}
+ { (d y^3)^2 + (d y^4)^2 \over 1 + X_1 + X_2 \cos^2 \alpha}
+ \sum_{i=5}^9( d x^i)^2 \Bigg \}\nonumber\\
A^{(4)} & = & - {X_2 \cos \alpha\, \sin \alpha \over 2} \Bigg \{ {1
\over 1+X}
+ {1 \over 1 + X_1 + X_2 \cos^2 \alpha} \Bigg \}
dt \wedge d y^1 \wedge d y^3 \wedge dy^4\nonumber\\
& & \qquad \mbox{}+\Bigg \{{1 -2X_1 \over 2 X_1}
+ {1 \over 2(1 + X_1 + X_2 \cos^2 \alpha)} - {1 + X_2 \over 2
X_1(1+X)}
\Bigg \} dt \wedge d y^2 \wedge d y^3 \wedge d y^4 \nonumber\\
& & \qquad\mbox{} + \ell_1^3 \sin^3 \theta\, \sin^2 \phi_1\,
\cos \phi_2\, d y^1 \wedge
d \theta \wedge d \phi_1 \wedge d \phi_3 \nonumber\\
& & \qquad\mbox{} + \ell_2^3 \cos \alpha\, \sin^3 \theta\,
\sin^2 \phi_1\, \cos \phi_2\,( \cos \alpha\, d y^1 - \sin \alpha\, d
y^2)
\wedge d \theta \wedge d \phi_1 \wedge d \phi_2 \nonumber\\
A^{(2)} & = & -{X_2 \sin \alpha \over 1+X }\, dt \wedge \bigg \{
\sin \alpha\,(1 + X_1) d y^1 + \cos \alpha\, d y^2 \bigg \}
\nonumber\\
B^{(b)} & = & -{X_2 \cos \alpha\,\sin \alpha \over 1 + X_1 + X_2
\cos^2\alpha}
d y^3 \wedge d y^4 \nonumber\\
e^{2 \phi_b} & = & {1 +X \over 1 + X_1 + X_2 \cos^2 \alpha}\, .
\labell{partialthree}
\eeqa
where we have also put $\vec x_1 = 0$ here for simplicity.
With $X_2=0$, we have a single D3-brane filling ($y^2,y^3,y^4$) and
delocalized in $y^1$. With $X_1=0$, one may verify that the result
describes a D(3,1)-brane bound state parallel to ($\sin\al\,y^1+
\cos\al\,y^2,y^3,y^4$) with the D1-branes lying in the first of
these
directions. Again the relative charge densities
of the bound state are determined by the rotation angle.

\section{Discussion} \labels{conc}

In this paper, we presented a new low-energy solution (\ref{newsol})
describing an arbitrary number $n$ of D-membranes oriented at angles
with
respect to one another.
We were also able to show that this configuration saturated the
BPS bound because the relative rotations between the membranes
are in an $SU(2)$ subgroup. As a result, the system preserves
one-quarter of the supersymmetries.

Our solution provides the most general supersymmetric
configuration containing (only) two D-membranes. One might
think of extending the rotations considered here to an arbitrary
$SU(2)$ rotation, but this generalization would only
change the overall orientation of our solution. Following
the analysis of \cite{Berkooz}, with three D-membranes one might
make $SU(3)$ rotations while still preserving one-eighth of the
supersymmetries. This would extend the space in which the rotations
act to produce an effective seven-dimensional world volume. It would
be interesting to find the corresponding background field solution.
For general $n$, one might consider $SU(n)$ rotations\cite{Berkooz},
however, in practice one would be limited to $SU(4)$ by the fact
that the spacetime is ten-dimensional.

By applying T-duality to the membrane solution (\ref{newsol}),
we produced solutions describing systems of higher dimensional
D-branes
oriented at angles, and also configurations involving
D($p$+1,$p$--1)-brane
bound states. Since supersymmetry is preserved by T-duality, these
other new solutions also preserve one-quarter of the
supersymmetries.
These configurations may be useful in trying to understand the
microscopic counting of states of new four-dimensional black holes
in Type II string theories.
By explicit construction, we have confirmed the existence
of a supersymmetric configuration including D0-branes and
D(4,2,2,0)-bound
states. These supersymmetric solutions were conjectured in
\cite{lifschytz},
where it was shown that the interaction potential precisely vanished
between these two objects.

After this research was carried out,
two new papers\cite{angledrama1,angledrama2}
appeared which discuss branes
oriented at angles in different contexts than considered in
the present paper. In \cite{angledrama1}, a construction is
presented
of a configuration of D4-branes tilted by a real $SO(2)$ rotation
and held in static equilibrium by the presence of D-membranes and
fundamental strings. In \cite{angledrama2}, a novel new
configuration
of angled D5-branes which preserve 3/16 of the supersymmetries.
At present there is no obvious connection between these
solutions and those presented here, however, it will be interesting
to explore this question in more detail.

\section*{Acknowledgments}
We gratefully acknowledge useful conversations with Ramzi Khuri
and Nourredine Hambli.
This research was supported by NSERC of Canada and Fonds FCAR du
Qu\'ebec.

\end{document}